# Evidence of Electromagnetic Absorption by Collective Modes in the Heavy Fermion Superconductor UBe$_{13}$

J. R. Feller, C.-C. Tsai, and J. B. Ketterson
*Department of Physics and Astronomy, Northwestern University, Evanston, Illinois 60208*

J. L. Smith
*Los Alamos National Laboratory, Los Alamos, New Mexico 87545*

and

Bimal K. Sarma
*Department of Physics, University of Wisconsin-Milwaukee, Milwaukee, Wisconsin 53201*

We present results of a microwave surface impedance study of the heavy fermion superconductor UBe$_{13}$. We clearly observe an absorption peak whose frequency- and temperature-dependence scales with the BCS gap function $\Delta(T)$. Resonant absorption into a collective mode, with energy approximately proportional to the superconducting gap, is proposed as a possible explanation. A one-parameter fit to the data provides a simple relation between $\Delta(T)$ and the collective mode energy.

The superconducting states in heavy fermion compounds [1] such as UPt$_3$ and UBe$_{13}$ exhibit behaviors that differ markedly from those predicted by the theory of Bardeen, Cooper, and Schrieffer (BCS). Quantities such as the specific heat and ultrasonic attenuation, for example, are enhanced at low temperatures, displaying a power-law dependence on $T$ instead of the usual exponential form. This suggests the existence of nodes in the superconducting energy gap. Even more surprising was the discovery of multiple superconducting phases in UPt$_3$. Such behavior has been explained, at least qualitatively, within the framework of Ginzburg-Landau theory, assuming a multi-component order parameter. Much of the work on heavy fermion superconductivity, especially from a theoretical point of view, has been strongly influenced by the paradigm provided by superfluid $^3$He. In the superfluid phases of $^3$He, quasiparticle pairs form in states with relative orbital angular momentum quantum number $l = 1$ ($p$-wave), as opposed to the BCS $l = 0$ ($s$-wave) state. Consequently, the order parameter possesses a large number of degrees of freedom. This in turn gives rise to multiple superfluid phases and a rich spectrum of order parameter collective modes [2, 3]. The observation and classification of collective modes, notably by means of ultrasound absorption experiments, was instrumental in determining the symmetries of the order parameters corresponding to each phase. With mounting evidence that the heavy fermion superconductors might be characterized by unconventional $^3$He-like order parameters, it was natural to wonder whether they too could support collective oscillations. A number of theoretical investigations [4-6] of unconventional charged superfluids, assuming order parameters of various symmetries, have predicted mode frequencies (measured relative to the energy gap) similar to those found in $^3$He. However, the extent to which damping, due in part to the presence of impurities (a complication not encountered in $^3$He work), should limit the observation of collective modes is a difficult theoretical problem that has not been adequately addressed. High frequency (~2 GHz) longitudinal ultrasound measurements [7] of UBe$_{13}$ revealed a sharp attenuation peak just below the superconducting transition temperature $T_c$. This was initially interpreted as the signature of a low-lying collective mode. It was soon after realized [8], however, that the attenuation peak could more simply and plausibly be explained in terms of enhanced pair-breaking and coherence effects, which are consequences of the unusually large quasiparticle effective masses. Similar features have since also been seen in UPt$_3$ [9] and URu$_2$Si$_2$ [10].

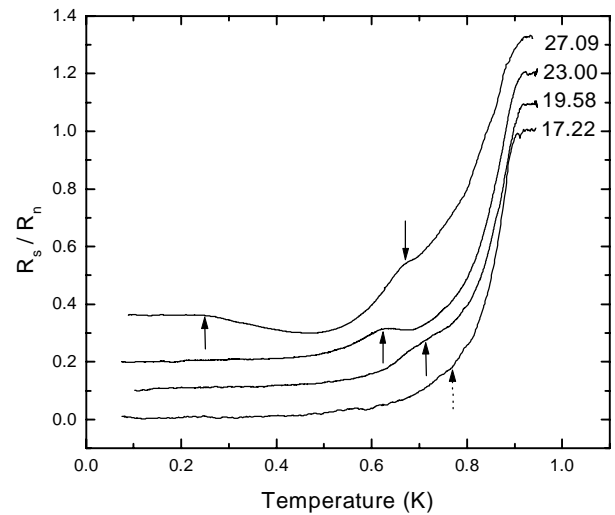

FIG. 1. Surface resistance of UBe$_{13}$ normalized to unity at $T_c$ and zero at $T \rightarrow 0$. Successive curves have been shifted upward for clarity. Each curve is labeled with the measurement frequency in GHz. Solid arrows indicate peak positions. The dotted arrow shows the position of a peak that is predicted but not clearly observed.

While very well suited to the study of $^3$He, ultrasound is not the ideal probe of collective mode behavior in heavy fermion superconductors. One expects characteristic energies on the order of the gap function $\Delta \sim k_BT_c$, which translates to a frequency of roughly 20 GHz. Ultrasonic measurements at such frequencies are problematic. Since we are dealing with a charged system, electromagnetic excitation of collective modes is a natural alternative. The obvious experiment is a microwave surface impedance measurement employing a resonant cavity. Low temperature microwave cavity measurements have been performed both on UPt$_3$ [11] and UBe$_{13}$ [12], but no clear evidence of collective mode absorption was found. While these null results are discouraging, they may merely be an indication of over-damping attributable to high impurity concentrations.

Here we report the results of recent surface impedance measurements of a very high quality UBe$_{13}$ single crystal at temperatures down to ~80 mK. The sample is an approximately rectangular slab with dimensions ~4.5 × 3.6 × 1.3 mm$^3$. It was prepared by putting U, Be, and Al in the atomic ratio 1:15:174 into an outgassed BeO crucible and heating to 1200°C in flowing helium gas. It was then cooled for 300 hours to the freezing point of aluminum. The aluminum was removed in a concentrated NaOH solution revealing UBe$_{13}$ single crystals with natural [100] facets of the cubic structure. The sample's surface was lightly etched in dilute sulfuric acid. For the measurements, a cylindrical, lead-plated copper cavity, mounted on a dilution refrigerator, was employed. The sample was top-loaded into the cavity through a central access hole. A frequency modulation technique was used, giving both the $Q$ and the resonant frequency $f_0$ as functions of temperature. From these quantities, the surface resistance and reactance were then calculated. We examined primarily the TE$_{01p}$ modes (TE$_{011}$, and its "overtones" TE$_{012}$, TE$_{013}$, and TE$_{014}$ with resonant frequencies at 17.22, 19.58, 23.00, and 27.09 GHz, respectively). These modes couple well with the antennae (inductive loops whose orientations are adjustable from the top of the cryostat) and so are easily distinguished from neighboring modes. They also seem to be "well-behaved" in that they give very reproducible results. In an ideal cylindrical cavity, TE$_{01p}$ is degenerate with TM$_{11p}$, but this degeneracy is lifted by perturbations in the top plate (e.g., introduction of the antennae), so that mode interference was not an issue. At the top of the cavity, where the sample was situated, the field distributions of all four modes are approximately the same. One caveat: since the sample sat at the center of the top plate (an unavoidable consequence of the top-loading feature) the fields were not distributed uniformly over its surface. Modes with $p > 4$ could not reliably be measured due, perhaps, to heavy losses in the coaxial lines, or to inefficient coupling.

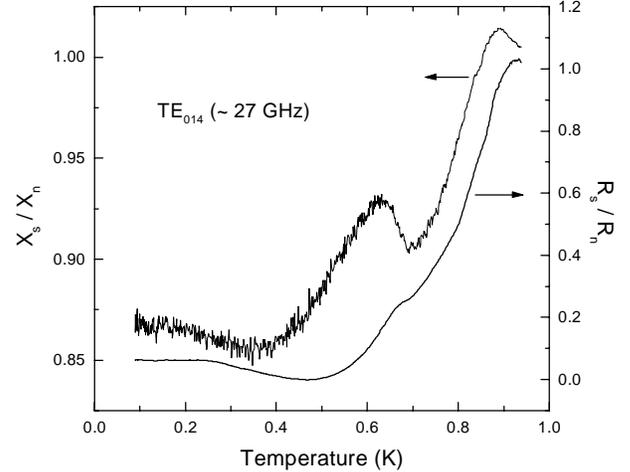

FIG. 2. The normalized surface resistance and reactance measured at 27.09 GHz.

The temperature dependence of the surface resistance for each measurement frequency is presented in Fig. 1. Each curve has been normalized to unity above $T_c = 0.905$ K, and zero at low temperatures. This is consistent with the Mattis-Bardeen theory [13], assuming the London limit. Here the actual normalization procedure is not critical, since we are primarily concerned with the frequency- and temperature-dependence of well-defined anomalies. Superimposed on the usual monotonic responses are prominent absorption peaks whose positions $T^*$ depend strongly on the measurement frequency. At the lowest frequency (17.22 GHz), no anomalies are seen before background subtraction. Above $T_c$, the surface impedance is flat, implying a very weak temperature dependence of the normal state resistivity. In Fig. 2 is shown both the normalized resistance and reactance for a single mode (TE$_{014}$ at 27.09 GHz). Each resistance peak is accompanied by an S-shaped anomaly in the reactance, its "zero-crossing" coinciding with the peak maximum. This is precisely the behavior one would expect at a resonance crossing. Moreover, the sense of the "S", indicating a transition from an "inductive" to a "capacitive" regime upon warming, implies that the slope of the resonance curve $\Omega(T)$ with respect to temperature is negative. Similar features are observed at lower frequencies; there they are less pronounced, but still well defined. The feature at $T_c$ is the well-known coherence peak.

In order to more accurately analyze the positions and shapes of the resistance anomalies, the monotonic backgrounds were subtracted. This was accomplished by first fitting the 17.22 GHz data with the normalized surface resistance predicted by the Mattis-Bardeen theory [13]. The only adjustable parameter was the ratio $c \equiv \Delta(0)/k_BT_c$, where $\Delta(0)$ is the value of the gap function at $T = 0$. Corresponding curves were then generated for the higher frequency modes using the same value of $c$. The physical validity of this fitting procedure will be addressed below. For the moment, it is merely used to provide smooth reference curves for the background subtractions. The positions of the 19.58 GHz and 23.00 GHz peak maxima can be read directly from the subtracted curves; they occur at reduced temperatures $t^* = T^*/T_c = 0.79$ and 0.69, respectively. At the highest frequency, 27.09 GHz, two anomalies are observed: a peak centered at $t = 0.74$, and a broad plateau at low temperatures. For the plateau, the value of $t^*$ (0.28) is taken as the midpoint of the corresponding reactance "S" (Fig. 2). This is somewhat ambiguous, but in the picture developed below, any $t^* < \sim 0.3$ would give similar results. Upon subtraction of the 17.22 GHz data, a broad feature below $T_c$ appears. It is nearly obscured by noise, but a prejudiced observer might interpret it as a washed-out peak at $t \sim 0.85$. This is indicated in Fig. 1 by the dotted arrow.

Our interpretation is that the resistance peaks appearing at 19.58 GHz and 23.00 GHz, along with the 27.09 GHz plateau, represent absorption by a single temperature-dependent resonant mode with frequency $\Omega_0(t)$. In this picture, the high temperature resistance peak seen at 27.09 GHz (represented by the single diamond in Fig. 3) is presumably indicative of a separate, higher frequency mode. As stated above, the reactance data implies that $d\Omega_0/dt < 0$ at each resonance crossing. The appearance of a plateau at high frequencies further implies that $d\Omega_0/dt \to 0$ as $t \to 0$. This temperature dependence is similar to that displayed by the BCS gap function $\Delta(t)$. If we make the assumption that

$$h\Omega_0(t) \approx b\Delta(t), \quad (1)$$

where b is a dimensionless constant, then one would expect to see an absorption (or resistance) peak at a temperature $t^*$ given by $hf = b\Delta(t^*)$, or

$$\frac{hf}{k_B T_c} = bc\delta(t^*), \quad (2)$$

where $\delta(t)$ is the gap function normalized to unity at $t = 0$, which is calculated from the usual BCS integral equation, and $c$ is the (unknown) scaling factor, defined above, that gives the value of $\Delta$ at absolute zero. The quantity $bc$ (this is the only fitting parameter) was adjusted so that the positions of the resistance peaks coincided with the curve $\delta(t)$. The resulting fit, with $bc = 1.442$, is shown in Fig. 3. The solid circles represent observed peaks. The open circle is the predicted position of the peak at $f = 17.22$ GHz, but for which there is no clear evidence in the data.

We propose that the observed anomalies are signatures of resonant absorption by an order parameter collective mode whose temperature dependence is approximately that of the order parameter itself, as expressed in Eq. (1). A number of such modes exist in superfluid $^3$He. The collective mode spectrum of a superconducting crystal is expected to be more limited, due to crystal field splitting. However, calculations assuming a p-wave ABM-like state [4] and d-wave $E_{1g}$ state with an order parameter of the form $k_x \pm ik_y$ [5, 6] yield an "optical" mode, analogous to the clapping mode found in $^3$He-A, with energy given by (1) with $b \sim 1.2$. Comparison of this with the value of the product bc determined above gives $c \sim 1.2$. Higher values of $c$ constrain the mode to lower frequencies relative to the energy gap. Lack of an independently determined value for this parameter hampers further analysis. The BCS theory gives $c = 1.76$, but for unconventional superconductors we should not be surprised to find significant deviations. Analysis of early UBe$_{13}$ specific heat data, assuming an ABM state and including strong-coupling corrections, yielded a value between 1.65 and 1.9 [14]. A naïve application of Mattis-Bardeen to the 17.22 GHz resistance curve, as described above, results in $c = 1.45$. This, however, is rather dubious considering that gap anisotropy has not been taken into account. Incorporating effects due to point nodes in an average way, we find that $c$ is increased to nearly the BCS value. In addition, it is known that impurities can act as pair-breakers, significantly affecting the electromagnetic response [15] (although we believe the impurity concentration to be quite small in our UBe$_{13}$ sample). A full analysis unfortunately introduces other unknown parameters. Traditionally, ultrasound has provided an effective probe of the energy gap. The attenuation of longitudinal ultrasound in UBe$_{13}$ [7] exhibits a clear $T^2$ dependence at low temperatures. Such behavior is typically taken as evidence for the existence of point nodes in the energy gap. A fit to the published data using a modification of the simple BCS expression for the attenuation coefficient [16], assuming an anisotropic energy gap of the form $\Delta(T,\theta) = \Delta_0(T)\sin\theta$ (resulting in point nodes at $\theta = 0, \pi$), reproduces the $T^2$ dependence and gives $c \sim 1.2$. Using this value, a near perfect correspondence between our data and the predicted collective mode energy is realized. It has, however, been demonstrated [8] that ultrasound measurements of unconventional superconductors, when approached in such a straightforward way, can be very misleading; it is not clear that one can even differentiate between different nodal structures (e.g., between point nodes and line nodes). Tunneling spectroscopy data might be less equivocal. Recent spectroscopy measurements [17] of $UBe_{13}$ suggest a much larger value of $c$, placing on it a *lower* limit of 3.35. This would imply a rather low-lying collective mode with energy $h\Omega_0 < 0.43\Delta$.

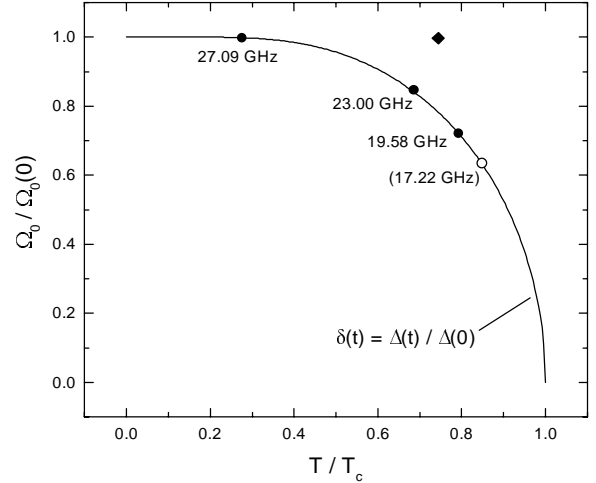

FIG. 3. Proposed collective mode frequency, normalized with respect to its value at absolute zero. It is assumed to be proportional to the gap function $\Delta(T)$. The solid circles represent the observed resistance peaks. The open circle is the predicted position of the peak at 17.22 GHz. The single diamond represents a second peak observed at 27.09 GHz.

It is expected that an order parameter collective oscillation will be damped both by quasiparticle excitations at the gap nodes and by the influence of impurities and other defects. The theory is not well developed at this time, but an overall decrease in damping effectiveness as $T \rightarrow 0$ is predicted [18]. This is qualitatively consistent with our observations. The width of the observed absorption peak can be taken as a measure of the damping strength. Each peak was fitted with a Lorentzian line shape defined by a temperature-dependent center frequency given by (1) and an effective relaxation time $\tau$, which is inversely proportional to the peak width (in the frequency domain). A physical interpretation of $\tau$ requires knowledge of the actual damping mechanisms, but it should depend strongly on the defect concentration, and therefore on the normal state carrier mean free time. The relaxation time is found to decrease dramatically as the measurement frequency $f_0$ is decreased (i.e., as $T^*$ approaches $T_c$): $\tau = 220 \cdot 10^{-12}$, $130 \cdot 10^{-12}$, and $70 \cdot 10^{-12}$ s at $f_0 = 27.09$, 23.00, and 19.58 GHz, respectively. For $f_0 = 17.22$ GHz, we have estimated a time ($\tau \sim 10 \cdot 10^{-12}$ s) that would render the resistive peak practically unobservable. We find, moreover, that if the relaxation times are decreased by an order of magnitude, the calculated resonance peaks are smeared beyond recognition. This points out the importance of high quality crystals.

To summarize, we have performed microwave surface impedance measurements of the heavy fermion superconductor $UBe_{13}$. Clearly seen is a resistance peak whose frequency- and temperature-dependence scales approximately as the BCS gap function. We interpret this as evidence of electromagnetic power absorption by an order parameter collective mode. This interpretation is consistent with existing theory. A one-parameter fit gives the mode energy at $T = 0$: $h\Omega_0(0) = 1.442 k_B T_c$. Further measurements in this vein on very pure samples of $UBe_{13}$ as well as other heavy fermion superconductors could serve to illuminate the still undetermined symmetries of the order parameters in these materials.

This work was supported by the National Science Foundation (NSF grants DMR-9120521, DMR-9309061, DMR-9704020, and DMR-9971123). Work at Los Alamos was performed under the auspices of the U. S. Department of Energy.